\documentclass[12pt]{article}
\usepackage{morefloats}
\usepackage{graphics}
\usepackage {epsfig}
\usepackage {ams}
\newfont{\rams}{msbm9 scaled\magstep1}
\newfont{\iams}{msbm9}
\newfont{\gotic}{eufm10 scaled\magstep1}
\newfont{\bellap}{eusm10 scaled\magstep1}
\newcommand{\rea}{\mbox{\rams \symbol{'122}}}

\newcommand{\rind}{\mbox{\iams \symbol{'122}}}

\newcommand{\nq}{{\sf n}_{q}}

\newcommand{\w}{\omega}
\newcommand{\hw}{\hbar \omega}

\newcommand{\kB}{k_{B}}
\newcommand{\TL}{T_{L}}
\newcommand{\kT}{\kB \TL}

\newcommand{\en}{\varepsilon}
\newcommand{\bk}{{\sf k}}
\newcommand{\xk}{{\sf x}}
\newcommand{\vg}{{\sf v}}
\newcommand{\Ef}{{\sf E}}

\newcommand{\bka}{\bk^{\prime}}
\newcommand{\kka}{(\bk, \bka)}

\newcommand{\tx}{(t, \xk)}

\newcommand{\enk}{\en(\bk)}
\newcommand{\enka}{\en(\bka)}

\newcommand{\ddm}{\hspace{1pt} \delta (\enka - \enk - \hw) }

\newcommand{\dmprime}{\hspace{1pt} \delta (\en^{\prime} - \en - \w) }
\newcommand{\ddp}{\hspace{1pt} \delta (\enka - \enk + \hw) }

\newcommand{\dpprime}{\hspace{1pt} \delta (\en^{\prime} - \en + \w) }

\newcommand{\ddi}{\hspace{1pt} \delta (\enka - \enk) }
\newcommand{\di}{\hspace{1pt} \delta (\en^{\prime} - \en) }

\newcommand{\Itre}{\int_{\scriptstyle \rind^{3}}}

\newcommand{\p}{\hspace{1pt} .}
\newcommand{\sv}{\hspace{1pt} ,}

\newcommand{\dm}{\displaystyle}
\begin{document}
\title{Moment inequalities and high-energy tails for electron distribution function of the
Boltzmann transport equation in semiconductors.}
\author{Orazio Muscato}
\date{}
\maketitle
\begin{center}
{Dipartimento di  Matematica e Informatica \\
Universit\`{a} di Catania \\
Viale Andrea Doria 6 - 95125 Catania, Italy }\\
\vspace{.5cm}

e-mail:{\sl  muscato@dmi.unict.it}
\end{center}
%
%
%
\begin{abstract}
\noindent In this paper we prove the existence of the high-energy
tails for electron distribution function of the Boltzmann equation
for semiconductors, in the stationary and homogeneous regime, in
the analytic band approximation and scattering with acoustic and
optical phonons and impurities.
We also prove numerically that the tail is a global maxwellian in
the parabolic band approximation, and in the quasi parabolic band
case, a power law of the global maxwellian.\\
%
%
%
\noindent
{\bf keywords} Boltzmann-Poisson system for semiconductors, Monte Carlo Method,
Semiconductors.
\end{abstract}
%
%
%
\section{Introduction}               %
%
%
%
In recent years, the dimensions of MOS devices have been scaled
down reaching submicrometric order. As a result, large electric
fields near the drain region generate hot or energetic electrons.
Hot electrons can heat the device with important consequences for
long-term reliability, or can be injected into the oxide creating
an instability in the device performance.
Moreover, injection of hot electron onto a floating gate has been
widely used as the writing mechanism in EPROM and flash EEPROM
device.
To model these effects is a difficult task, because they depend
upon the small fraction of carriers with energy above certain
thresholds and hence the knowledge of the tail of the electron
energy distribution (hereafter EED) is required.

\noindent The natural framework for describing these regimes is
the Boltzmann Transport Equation (BTE) with the adequate
scattering mechanisms, coupled with the Poisson equation.
To solve the BTE is an hard task: recently deterministic solvers
of the BTE, based of finite difference schemes, have been
introduced \cite{carrillo} but their computational efforts is
comparable to that of the Direct Simulation Monte Carlo.

\noindent Another alterative is to use Extended hydrodynamic
models, based on the higher order moments of the BTE \cite{AnRo1},
but their sensitivity to the tail of the EED is an open question.

\noindent At the moment, the Monte Carlo (MC) solution of the BTE
represents the most accurate approach for describing the tail of
the EED, because the correct scattering mechanisms as well as the
bands structure is taken into account. Unfortunately, the MC
approach is CPU consuming, hence possibly unsuitable for extensive
use in technology development.
What we know from MC simulations is that the EED shape, for high
energies, is maxwellian \cite{abramo,childs} (for parabolic bands
and acoustic and optical phonons scattering), whereas the
electron-electron interaction modifies this asymptotic behaviour
enhancing the tail \cite{abramo,fischetti,chung}.

\noindent The understanding of these phenomena has been the
subject of a wide debate: according to some authors  such  tails
reflect the distribution at the device boundaries due to ballistic
transport \cite{lacaita,mastropasqua}, or has been caused by
'superlucky' electrons which have absorbed more phonons than they
have emitted \cite{abramo}.

\noindent A lot of effort has also been put into the development
of analytical expression for the whole EED.
Heuristic models for the EED are based on expansions around a
global maxwellian \cite{henning} , or polynomial in the
exponential function \cite{AnRo1,cassi}, or a combination of a
global maxwellian and a power law of a global maxwellian (also
known as  'stretched exponential') \cite{grasser} in order to
describe 'cold' and 'hot' electron populations.
The parameters appearing in these EEDs can be determined by
considering some approximation of the BTE and by fitting the
resulting model to MC data.
%
%
%
%

\noindent The problem of the description of the high-energy tails
has been tackled in a different context as in kinetic models for
granular flows.
A variety of analytical techniques have been developed such as by
comparing the 'gain' and 'loss' terms in the BTE \cite{brito}, or
by means of a rigorous pointwise lower estimate \cite{gamba}, or
by means of $L^1$ weighted tail control \cite{bobylev}, or by
using the so-called inelastic Maxwell models
\cite{boby_ga_carri,boby_cerci}.

\noindent The aim of this paper is to tackle the high-energy tail
problem rigorously in the BTE framework, by using an analytical
technique adapted from granular media which is  based on moments
estimate \cite{bobylev}. In this way, we are able to prove the
existence of the EED tail which has the form of a 'stretched
exponential'.
The paper is organized as follows. In Section 2 the basic
equations are written. In Section 3 the 'stretched exponential'
for the tail is introduced and we formulate the main results.
Section 4 is devoted to the moment inequalities. In Section 5 we
present the proof of the main theorem. In Section 6 simulation
results obtained by means of Direct Simulation Monte Carlo are
shown and conclusion are drown in Section 7.
%
%
%
%
\section{Basic Equations}
%
%
%
The Boltzmann transport equation (BTE) for one conduction band
\cite{Markowich,JacoboniReggiani} is
\begin{eqnarray}
\frac{\partial f}{\partial t}+\vg(\bk)\cdot\nabla_{\xk}{f}
-\frac{q}{\hbar}\,\Ef\cdot\nabla_{\bk} {f}= {\cal Q}[f] \quad ,
\label{Boltzmann}
\end{eqnarray}
where the unknown function $f(t,\xk, \bk)$ represents the
probability density of finding an electron at time $t$ in the
position $\xk \equiv (x_1,x_2,x_3)$ with the wave-vector $\bk
\equiv (k_1,k_2,k_3)$, and $q$ is the absolute value of the
electron charge.
The domain of $\bk$ is the first Brillouin zone ${\cal B}$.
\noindent In the neighborhood of the band minimum the dispersion
relation can be considered approximately  {\it quasi-parabolic}:
\begin{eqnarray}
      \en(\bk) \left[ 1 + \alpha \en(\bk) \right] = \gamma =
      \frac{ \hbar^2 \bk^2}{2 m^\star},  \label{Kane}
\end{eqnarray}
where $m^\star$ denotes the effective electron mass, which is 0.32
$m_e$ (free electron mass) in silicon, $\alpha$ the non
parabolicity factor (0.5 $eV^{-1}$),and $\hbar$ the Planck
constant divided by $2\pi$.
With this approximation the Brillouin zone ${\cal B}$ coincides
with $\rea^3$.

\noindent From eq.(\ref{Kane}), the electron group velocity $\vg
\equiv (v_1,v_2,v_3)$ writes
\begin{eqnarray}
\vg(\bk)=\frac{1}{\hbar}\nabla_{\bk}\en = \frac{\hbar
\bk}{m^\star} \frac{1} {\sqrt{1 + 4 \alpha \gamma}}\quad .
\label{vel}
\end{eqnarray}
The parabolic band approximation is obtained from the previous
equations with $\alpha$=0.

\noindent The electric field $\Ef\tx \equiv (E_1,E_2,E_3)$
satisfies the Poisson equation
\begin{eqnarray}
\Delta (\epsilon \phi) = q\left[ n\tx - N_D(\xk) + N_A(\xk)
\right] \nonumber
\end{eqnarray}
\begin{eqnarray}
\Ef=-\nabla_{\xk}{\phi}   \nonumber
\end{eqnarray}
where $\phi \tx $ is the electric potential, $N_D$ and $N_A$,
respectively, are the donor and acceptor densities, $\epsilon$ the
dielectric constant, $n$ the electron density given by
\begin{eqnarray}
n\tx=\Itre{f(t,\xk,\bk)} \p \nonumber
\end{eqnarray}
In general, for low electron density, the collision operator can
be schematically written as
\begin{eqnarray}
{\cal Q}[f] = \Itre{\left[ w(\bka,\bk) f(\bka) -
  w(\bk,\bka) f(\bk)\right]} d\bka   \label{collisione}
\end{eqnarray}
where the first term represents the gain and the second one the
loss, and $w\kka$ is the transition probability per unit time from
a state $\bk$ to a state $\bka$.

\noindent The main scattering mechanisms in a silicon
semiconductor are the electron-phonon interactions, the
interaction with impurities, the electron-electron scatterings and
the interaction with stationary imperfections of the crystal, as
vacancies.

\noindent In this paper we shall consider electron-phonon
interactions, and interaction with the impurities
\cite{JacoboniReggiani}, i.e.
\begin{eqnarray}
w\kka & =& w_{ac}\kka + w_{npo}\kka + w_{imp}\kka \label{kernel}
\p
\end{eqnarray}
The first term represents the acoustic phonon scattering in the
elastic approximation
\begin{eqnarray*}
w_{ac}\kka &=& K_{0}\quad \ddi  \\
K_{0} &=& \frac{\kT \Xi_{d}^{2}}{4\, \pi^{2}\hbar\,\rho \,
v_{s}^{2}} \label{ker} \sv
\end{eqnarray*}
where $T_L$ is the lattice temperature, $\Xi_d$ the acoustic
phonon deformation potential, $\rho$ the silicon mass density and
$v_s$ the sound velocity of the longitudinal acoustic mode;
the second term represents the non-polar optical scattering
\begin{eqnarray*}
w_{npo}\kka & =& K_{1} \left[ \nq \ddm +
(\nq + 1 )\ddp \right] \\
K_{1} &=& \frac{(D_t K)^{2}}{8\, \pi^{2}\rho\,
 \omega}  \label{keru}
\end{eqnarray*}
where $\hw$ is the optical phonon energy and $\nq$ the phonon
equilibrium distribution which, according to the Bose-Einstein
statistics, is given by
$$
\nq = \frac{1}{\exp (\hbar \omega/k_B T_L) -1} \quad .
$$
The third and last term represents the ionized impurity scattering
\begin{eqnarray*}
w_{imp}\kka &=&  \frac{K_{2}}{\left[\beta^2 + 2|\bk|^2(1-\cos
\theta)\right]^2}\quad  \ddi
\\[0.5cm]
K_{2} &=& \frac{4 Z^2 n_I q^4}{\hbar \kappa^2}\\[0.5 cm]
\cos \theta & = & \frac{\bk \cdot \bka}{|\bk| |\bka|}
\end{eqnarray*}
The values of the coupling constants and the other parameters used
for silicon are given in Table I.

The electron-electron interaction is taken into account in the
framework of the mean field approximation through the Poisson
equation. This is reasonable since we consider the case of low
electron density and, therefore, we can neglect the short range
collisions between electrons.
%
%
%
\section{Exponential tails}
%
%
In the following we shall draw our attention to the asymptotic
behaviour of the steady state distribution function of the BTE
$$
f_{\infty}(\bk) = \lim_{t \rightarrow \infty} f(t,\bk) \nonumber
$$
for large $|\bk|$, i.e. the high-energy tails.
We expect that the  behaviour of this solution is given by the so
called 'stretched exponential', i.e.
\begin{eqnarray}
f_{\infty}(\bk) \simeq  \exp\left[-r \en^{s}\right] \quad , \quad
|\bk| \rightarrow \infty \label{f_distr}
\end{eqnarray}
where $r$ and $s$ are some positive constants.
We introduce the functionals \cite{bobylev}
\begin{eqnarray}
{\cal F}_{r,s}(f) = \Itre f(\bk) \exp\left[r \en^{s}\right] d\bk
\label{fun}
\end{eqnarray}
and study the values of $s$ and $r$ for which these functionals
are positive and finite.
We can give the following definition:
\newpage
{\bf Definition}. We say that the function $f$ has an {\it
exponential tail of order s} $>$ 0, if the following supremum
\begin{eqnarray}
r^\star_s = \mbox{sup} \{ r>0 \quad |  \quad {\cal F}_{r,s}(f) < +
\infty \} \label{r}
\end{eqnarray}
is positive and finite.\\[0.5cm]

\noindent The functionals (\ref{fun}) can be represented by using
the {\it symmetric moments} of the distribution function, i.e.
\begin{eqnarray}
m_p = \Itre f(\bk) \en^{p} d\bk \quad , \quad p \in \rea_+ \quad .
\label{sym_moments}
\end{eqnarray}
In fact by expanding the exponential function in (\ref{fun}) into
Taylor series we obtain (formally):
\begin{eqnarray}
{\cal F}_{r,s}(f) = \Itre f(\bk)\left(\sum_{n=0}^\infty
\frac{r^n}{n!} \en^{sn} \right) d\bk = \sum_{n=0}^\infty
\frac{m_{sn}}{n!}r^n  \quad . \label{taylor}
\end{eqnarray}
In order the expansion (\ref{taylor}) have a sense, we shall
suppose that the moments of all orders are finite.
Then the value $r^\star_s$ can be interpreted as the {\bf radius
of convergence} of the series (\ref{taylor}), and the order of the
tail $s$ is therefore the value for which the series has a
positive and finite radius of convergence.
For investigating the summability of the series (\ref{taylor}) we
look for estimates of the sequence of moments $\{m_p\}$, with $p =
sn$, $n=0,1,2..$ , and study the dependence of the estimate on
$s>0$. We shall be interested in the situation when the sequence
of the coefficients satisfies
\begin{eqnarray}
\frac{m_{sn}}{n!} \le C Q^n , \quad n=0,1,2... \label{stima}
\end{eqnarray}
where $C, Q$ are  positive constants. In fact, according to the
Hadamard theorem
\begin{eqnarray*}
\lim_{n \rightarrow \infty} \sqrt[n]{\frac{m_{sn}}{n!}}=
\frac{1}{r^\star_s}
\end{eqnarray*}
and from the estimate (\ref{stima}) we have
\begin{eqnarray}
r^\star_s \ge \frac{1}{Q} > 0 \quad . \label{raggio}
\end{eqnarray}
To achieve the estimate (\ref{stima}), we shall study the moments
equations obtained by integrating against $\en^{p}$ the BTE, in
the {\bf stationary and homogeneous} regime :
\begin{eqnarray}
{\cal Q}_p + G_p = 0  \label{steady}
\end{eqnarray}
where
\begin{eqnarray}
{\cal Q}_p &=& \Itre  {\cal Q}[f] \en^{p} d\bk  \label{Q_p} \\
 G_p &=& \frac{q}{\hbar}\Itre \,\Ef\cdot\nabla_{\bk}f \en^{p}
d\bk \label{G_p}
\end{eqnarray}
We underline that in the stationary and homogeneous regime, the
electric field $\Ef$ is a constant.
Now we can formulate the main result of this paper: \\[0.3cm]
{\bf Theorem 1}. {\tt Let $f(\bk)$ be a nonnegative solution of
the BTE (\ref{Boltzmann}) in the sta-\\tionary and homogeneous
regime, that has finite moments of all orders.\\
Then the supremum $r^\star_s$ defined in (\ref{raggio}) is finite
for some s$>0$,and in the energy range}
$\hw < \en < \en_0 $, $\quad \forall \en_0>0$.  \\[0.3cm]

\noindent To the best of our knowledge this is the first theorem
which proves the existence of energy tails for the BTE in the
semiconductor framework.
This theorem proves also that the 'stretched exponential'
(\ref{f_distr}), used in an heuristic way to model the EED tail,
has a theoretical background.
On the other hand, by using this technique, no information is
given about the order of the tail $s$. This order will be
determined by MC simulations in sec. 6.
Our result is similar to that found in \cite{bobylev} for the BTE
for granular media, where a shear flow model is considered as
forcing term. But the scattering mechanism for granular flows is
completely different respect to the semiconductor case, because
binary inelastic collisions are considered between perfectly
smooth hard sphere granular particles. The only analogy between
the two cases, is the presence of an inelastic scattering
mechanism in the collisional operator.
%
%
%
\section{Moment inequalities}
%
%
Let us introduce non dimensional variables:
\begin{eqnarray*}
\tilde{k} = \frac{k}{k_0}\quad , \quad \tilde{v} = \frac{v}{v_0}
\quad , \quad \tilde{\en} = \frac{\en}{\en_0} \quad , \quad
\tilde{E} = \frac{E}{E_0} \quad , \quad \tilde{\omega}=
\frac{\hbar \omega}{\en_0}, \quad \tilde{\alpha}=\alpha k_B T_L
\end{eqnarray*}
where
\begin{eqnarray*}
k_0 = \frac{\sqrt{m^\star k_B T_L}}{\hbar} \quad , v_0=
\sqrt{\frac{k_B T_L }{m^\star}} \quad , \en_0 = k_B T_L \quad ,
\quad E_0 = \frac{k_B T_L \sqrt{m^\star k_B T_L}}{q\hbar}
\end{eqnarray*}
consequently by omitting the $\tilde{}$, we can write
$$
k^2 = 2 \en (1 + \alpha \en), \quad \vg = \frac{\bk}{1 + 2\alpha
\en}, \quad \en = \frac{-1 + \sqrt{1 + 2\alpha k^2}}{2 \alpha} \p
$$

\noindent Now we change the variable $\bk$ into spherical
coordinates, i.e.
\begin{eqnarray}
\bk \equiv(k_x, k_y, k_z) \longrightarrow (k\, \sin \theta \cos
\phi, k\, \sin \theta \sin \phi, k\, \cos \theta) \label{sferico}
\end{eqnarray}
and the density-of-state and the solid angle over the unit sphere
are respectively:
\begin{eqnarray*}
J = (1 + 2\alpha \en) \sqrt{2\en(1+\alpha \en)}, \quad d\sigma =
\sin \theta \,d\theta \, d\phi \, \p
\end{eqnarray*}
The symmetric moments (\ref{sym_moments}) write:
\begin{eqnarray}
m_p = \Itre f(\bk) \en^{p} J \, d\en \, d\sigma \p \label{m_p}
\end{eqnarray}
Under suitable conditions on smoothness and decay for $\en$ large
of the solutions $f(\bk)$, the eq.(\ref{G_p}) writes:
$$
G_p = -p |\Ef| \Itre f(\bk) \, \en^{p} \frac{\Ef \cdot \bk}{|\Ef|
|\bk|} \frac{\sqrt{2\en(1+\alpha \en)}}{\en (1 + 2\alpha \en)}\,
d\bk \, ,
$$
and from the definition of scalar product, and the eq.(\ref{m_p})
we have the following bounds :
\begin{eqnarray}
-2p |\Ef| m_{p - \frac12} \le G_p \le 2p |\Ef| m_{p - \frac12} \,
\label{dis_gp} \p
\end{eqnarray}
Now we manipulate the term ${\cal Q}_p$ (\ref{Q_p}):
\begin{eqnarray}
{\cal Q}_p = \Itre {\left[ w(\bka,\bk) f(\bka) -
  w(\bk,\bka) f(\bk)\right]} \en^{p} d\bka d\bk =
\Itre w(\bk,\bka) f(\bk)\left[(\en^{\prime})^{p} - \en^{p}
\right]d\bka d\bk \nonumber
\end{eqnarray}
and from the scattering rate (\ref{kernel}) we have:
\begin{eqnarray}
{\cal Q}_p &=& {\cal Q}_p^{(ac)} + {\cal Q}_p^{(npo)} + {\cal Q}_p^{(imp)} \nonumber \\
{\cal Q}_p^{(ac)} &=& K_0 \Itre \di f(\bk) \left[(\en^\prime)^{p}
-(\en)^{p} \right]d\bka d\bk \nonumber \\
{\cal Q}_p^{(npo)} &=& K_1 \Itre f(\bk) \left[\nq \dmprime + (\nq
+1) \dpprime \right]\left[(\en^\prime)^{p}-(\en)^{p}\right] d\bka
d\bk \nonumber \\
{\cal Q}_p^{(imp)} &=& K_2 \Itre \di f(\bk) \left[(\en^\prime)^{p}
-(\en)^{p} \right]d\bka d\bk \nonumber
\end{eqnarray}
The terms ${\cal Q}_p^{(el)}$ and ${\cal Q}_p^{(imp)}$ vanish due
to the presence of $\di$. By changing variables into spherical
ones, after some manipulation, we have:
\begin{eqnarray}
{\cal Q}_p = {\cal Q}_p^{(npo)}={\cal Q}_p^{(+)} + {\cal
Q}_p^{(-)} \nonumber
\end{eqnarray}
where
\begin{eqnarray}
{\cal Q}_p^{(+)} =&& 4\pi 2^{\frac12}\nq K_1 \Itre f(\bk)
\left[(\en + \w)^p - \en^p \right]\left[1+2\alpha(\en +
\w)\right] \nonumber \\
&&\times \sqrt{(\en + \w)\left[1 + \alpha(\en+\w)\right]} \,J d\en
\, d\sigma \label{Qp_p}
\end{eqnarray}
\begin{eqnarray}
{\cal Q}_p^{(-)} = && 4\pi 2^{\frac12}(\nq +1) K_1 \Itre f(\bk)
\left[(\en - \w)^p - \en^p \right]\left[1+2\alpha(\en -
\w)\right] \nonumber \\
&&\times \sqrt{(\en - \w)\left[1 + \alpha(\en - \w)\right]}
\,\Theta(\en - \w) J d\en \, d\sigma \label{Qp_m} \p
\end{eqnarray}
Now we give an upper bound for ${\cal Q}_p^{(+)}$. By using the
following inequalities
\begin{eqnarray*}
\sqrt{\en + \w } &<& \en + 2\w \\
(\en + \w)^p & \le & 2^{|p-1|} \left[\en^p + \w^p \right] \quad ,
\forall p \ge 0
\end{eqnarray*}
(where $\omega \simeq$ 2.4369 in non dimensional units) and the
definition (\ref{m_p}), we obtain:
\begin{eqnarray}
{\cal Q}_p^{(+)} \le && 2^{\frac32}\pi \nq K_1 2^{|p-1|}
 \left\{\frac{\alpha^2}{2}m_{p+3} + \frac12(3\alpha + 8\alpha^2 \w)
m_{p+2} + (10\alpha^2 \w + 9\alpha \w+1) m_{p+1} \right. \nonumber \\
&& + 4 \w(2\alpha^2 \w^2 + 3\alpha\w +1) m_p +
2^{p-1}\alpha^2\w^{p+1}m_3 + 2^{p-1}\w^{p}(3\alpha +
8\alpha^2\w)m_2 \nonumber \\
&& \left. + (2\w)^p(10\alpha^2 \w^2 + 9\alpha \w +1)m_1 +
\w^{p+1}2^{p+2}(2\alpha^2\w^2 + 3\alpha\w +1) m_0 \right\}
\label{dis_qp} \p
\end{eqnarray}
Now we give an upper bound for ${\cal Q}_p^{(-)}$. By using the
inequalities:
\begin{eqnarray*}
&& \sqrt{\en - \w} > a (\en -\w)^3 \quad, \mbox{for} \quad \w < \en <\w + \frac{1}{a^{\frac{2}{5}}}
\quad \forall a>0 \\
&& \en^p - (\en - \w)^p > \en^{p-1}, \mbox{for} \quad \en > \w
\quad \mbox{and} \quad p \ge \frac{1}{\w}\simeq 0.41
\end{eqnarray*}
we can write
\begin{eqnarray*}
\left[ (\en - \w)^p -\en^p \right] \sqrt{\en - \w}< -\en^{p-1} a
(\en -\w)^3
\end{eqnarray*}
and ${\cal Q}_p^{(-)}$ can be estimated as
\begin{eqnarray*}
{\cal Q}_p^{(-)} \le && -4\pi 2^{\frac12}(\nq +1) K_1 a\Itre
f(\bk) \en^{p-1}(\en - \w)^3 \left[1+2\alpha(\en - \w)\right]
\nonumber \\
&& \times \sqrt{1+\alpha(\en - \w)} \Theta(\en - \w) J d\en \,
d\sigma  \p
\end{eqnarray*}
By using the further inequalities
\begin{eqnarray*}
&& \sqrt{1+ \alpha(\en - \w) } > \beta \en ,\quad \mbox{with} \,
\en < \frac{\alpha + \sqrt{\alpha^2 + 4\beta^2(1-\alpha
\omega})}{2\beta^2} \quad , \forall \beta >0 \\
&& 1+ 2\alpha(\en - \w) > \Gamma \en , \quad \mbox{with} \, \en <
\frac{1-2\alpha \omega}{\Gamma -2\alpha} \quad , \forall \Gamma
> 2\alpha  \label{dis2}
\end{eqnarray*}
we obtain the final upper bound for ${\cal Q}_p^{(-)}$
\begin{eqnarray}
{\cal Q}_p^{(-)} \le -\pi (\nq+1) K_1 a \beta \Gamma
\left\{2^{-\frac23} m_{p+4} - 3\w 2^{-\frac12} m_{p+3} + 3\w^2
2^{\frac12} m_{p+2} + \w^3 2^{\frac32} m_{p+1} \right\}
\label{dis_qm} \p
\end{eqnarray}
Now we are ready to obtain a recursive inequality for the moments.
From eq.(\ref{steady}) and the inequality (\ref{dis_gp}), we get
\begin{eqnarray*}
-2p |\Ef| m_{p - \frac12} \le {\cal Q}_p = {\cal Q}_p^+ + {\cal
Q}_p^-
\end{eqnarray*}
and using the upper bounds eqs.(\ref{dis_qp}),(\ref{dis_qm}), we
obtain finally
\begin{eqnarray}
\left\{
\begin{array}{l}
m_{p+1} \le R 2^{|p-4|}m_{p} + S_2 2^{|p-4|}m_{p-1} + S_1
2^{|p-4|}m_{p-2} +  S_0 2^{|p-4|}m_{p-3} +  T (p-3) m_{p-\frac72} +\\[0.8 cm]
V_3 2^{|p-4|}2^{p-4} \w^{p-2}m_3 + V_2 2^{|p-4|}2^{p-4}
\w^{p-3}m_2
+ V_1 2^{|p-4|}2^{p-3} \w^{p-3}m_1 + \\[0.8 cm]
V_0 2^{|p-4|}2^{p-1} \w^{p-2}m_0  \\[0.8cm]

\forall p \ge 3+ \frac{1}{\dm \w} \simeq 3.4104, \quad \w < \en <
\en_{M}
\end{array} \label{dis_mom}
\right.
\end{eqnarray}
where $\en_{M}$ is an arbitrary fixed energy value, and the
coefficients are positive constants given in the Appendix.

\noindent Assuming some properties of the moments of lower order,
we can use the recursive inequality (\ref{dis_mom}) to obtain
information about the behaviour of the moments $m_p$, for $p$
large.
%
%
%
\section{Proof of theorem 1}           %
%
%
We want to prove that the coefficients of the series
(\ref{taylor}) satisfy the inequality (\ref{stima}).
First of all we show that {\bf for every} $p\ge 1$ there exist $C$
and $Q$ positive constants, such that
\begin{eqnarray}
m_{p} \le C Q^{p}  \label{main_inequa} \quad .
\end{eqnarray}

\noindent To obtain the desired estimate, we split our proof in
two parts
\begin{enumerate}
\item We fix $p_0$ and we prove that the inequality (\ref{main_inequa}) holds for $1 \le p \le
p_0$. \\
We introduce the Jensen's inequality \cite{cercignani}: \\[0.5cm]
{\it  Let be ${\cal C}[g(\bk)]$ a concave function and $f(\bk)$ a
density probability, then
\begin{eqnarray}
\Itre f(\bk) {\cal C}[g] d\bk \le {\cal C} \left[ \Itre f(\bk)
g(\bk) d \bk \right] \label{jensen} \p
\end{eqnarray}
}\\[0.5cm]
For 1$\le p\le p_0$  we choose:
$$
g(\bk)= \en^{p_0} , \quad {\cal C}[g]=g^{\frac{p}{p_0}}
$$
and from (\ref{jensen}) we obtain
$$
\Itre f(\bk)\left[\en^{p_0}\right]^{\frac{p}{p_0}} d\bk \le \left(
\Itre f(\bk) \en^{p_0} d\bk \right)^{\frac{p}{p_0}}
$$
which can be written as:
\begin{eqnarray}
\left\{
\begin{array}{l}
m_p \le (m_{p_0})^{\frac{p}{p_0}} = C Q^p \label{dis_mom1}
\\[0.8cm]
1 \le p \le p_0
\end{array}
\right.
\end{eqnarray}
with $C$=1 and $Q$=max$\{1,\left(m_{p_0}\right)^\frac{1}{p_0}\}$.
\item We prove the inequality (\ref{main_inequa}) for $p_0<p<p_0 +1$. \\
We can estimate the moments
$m_{p_0-1}$,$m_{p_0-2}$,$m_{p_0-3}$,$m_{p_0- \frac72}$ which
appear in the left-hand- side of (\ref{dis_mom}) by using the
inequality (\ref{dis_mom1}), supposing  $p = p_0 \ge \frac{13}{2}
+ \frac{1}{\omega}$.
The eq.(\ref{dis_mom}) writes:
\begin{eqnarray}
m_{p_0 +1} \le (2Q)^{p_0} \Gamma_1 + (4\omega)^{p_0}\Gamma_2
\label{dis_mom2}
\end{eqnarray}
where
\begin{eqnarray}
\left\{
\begin{array}{l}
\Gamma_1 = \frac{\dm 1}{\dm 2^4} \left[ R + S_2 Q^{-1} + S_1Q^{-2}
+ S_0 Q^{-3} +
4 T  Q^{-\frac72} \right] > 0 \\[0.8cm]
\Gamma_2 = \frac{\dm 1}{\dm 2^8 \omega^2}V_3 m_3 + \frac{\dm
1}{\dm  2^8 \omega^3}V_2 m_2 + \frac{\dm 1}{\dm 2^7 \omega^2}V_1
m_1 + \frac{\dm 1}{\dm  2^5\omega^2}V_0 m_0 >0 \\[0.8cm]
\end{array} \nonumber
\right.
\end{eqnarray}
Now eq.(\ref{dis_mom2}) can be written as
\begin{eqnarray}
m_{p_0 +1} \le \tilde{\Gamma} \tilde{Q}^{p_0} \label{dis_mom4}
\end{eqnarray}
where $\tilde{\Gamma}$ = 2 max($\Gamma_1,\Gamma_2$), and
$\tilde{Q}$ =  max(2Q,4$\omega$)

By using induction arguments, from eq.(\ref{dis_mom4}) we can
conclude that the inequality (\ref{main_inequa}) holds for
$p_0<p<p_0 +1$.
\end{enumerate}

\noindent For every $p \ge$ 1 we can cover the whole interval by
using the previous arguments.
A similar theorem is valid in the parabolic band approximation.
%
%
\section{Simulation results}           %
%
In the following we shall consider the bulk, where an homogeneous
electric field is frozen into an homogeneous piece of silicon.
The experimental evidence of the high-energy tails of the EED has
been widely proved in literature  by MC simulations.
In the case of analytic parabolic band approximation (with
acoustic and optical phonons scattering) this tail is a global
maxwellian \cite{abramo,childs}.
So in principle we know, in this case, that the order of the tail
in eq.(\ref{f_distr}) is $s=1$.
What is unknown is the radius of convergence $r_s^\star$ of the
series (\ref{taylor}), which is the inverse of the maxwellian
temperature. This parameter will be determined by means of MC
simulations.

\noindent Several fitting formula are given in literature for the
whole EED in the quasi parabolic band approximation.
Here we shall check the EED tail by means of eq.(\ref{f_distr})
determining, by using MC simulations, $r_s^\star$ and $s$ in the
quasi parabolic band approximations.

\noindent An histogram of the numerical steady solution is
obtained by changing variable  from $\bk$ to $(\en, \theta, \phi)$
according to (\ref{sferico}), discretizing the whole energy space
in a system of concentric shells with increasing radius
\begin{eqnarray*}
\rho_n = n \, h , \quad  n=1,...,N , \quad h = \frac{R}{N}
\end{eqnarray*}
and by counting the number of particles which are in the
corresponding shells
\begin{eqnarray*}
f_1 &=& \{ \# \, \mbox{particles}  : \en < \rho_1 \} \\
f_n &=& \{ \# \, \mbox{particles}  : \rho_{n-1} \le \en < \rho_n \} \quad  n= 2,...,N \\
f_{N+1} &=&  \{ \# \, \mbox{particles}  : R \le \en \} \quad \p
\end{eqnarray*}
In order to obtain the parameters in (\ref{f_distr}), we rewrite
this equation into the form:
\begin{eqnarray*}
f_n = c \exp\left[-\left(\frac{\en_n - \en_{n_0}}{T_M}\right)^{s}
\right] \quad, n\ge n_0
\end{eqnarray*}
and plot the pairs $(x_n, y_n)$
\begin{eqnarray*}
x_n = \ln(\en_n - \en_{n_0}) \quad , y_n=\ln(\ln f_{n_{0}} - \ln
f_n) \quad , n=n_0+1,...,N   \quad \p
\end{eqnarray*}
Thus we expect that
\begin{eqnarray}
y_n = s x_n + b \label{fit}
\end{eqnarray}
where
\begin{eqnarray}
b = - s \ln T_M
\end{eqnarray}
is an almost linear plot for large values of the energy, whose
slope is the exponent $s$, and the y-intercept is related to the
parameter $T_M$ (whose dimension is an energy).

\noindent In figure 1 we show the simulation results obtained in
the parabolic band approximation (i.e. $\alpha$=0 in
eqs.(\ref{Kane}),(\ref{vel})) with an electric field of 80 KV/cm.
The solid curve shows the MC data $y_n$, while the dashed straight
line represents eq.(\ref{fit}) with $s$=1 and $T_M$= 4907 K.
Thus the asymptotic EED is the global maxwellian
\begin{eqnarray*}
f_{\infty} \simeq \exp\left[-\frac{\en}{T_M}\right] \p
\end{eqnarray*}

\noindent In figure 2 we show the slope $s$ as function of the
electric field, both for parabolic and quasi parabolic band
approximations. In the parabolic band approximation this value is
approximately one (i.e. the EED has a global maxwellian tail), but
in the quasi parabolic case the EED tail is given by the
'stretched exponential' (\ref{f_distr}). The parameter $s$ is a
function of the electric field, and for small field values,
approaches the value one, i.e. the tail is a global maxwellian.

\noindent In figure 3 we show the parameter $T_M$ as function of
the electric field, both for parabolic and quasi parabolic band
approximations. For moderate values of the electric field this
parameter coincides with the lattice temperature (dashed straight
line), but for higher values this parameter increases.
\noindent This behaviour can be justified as follows: $r^\star_s$
is the convergence radius of the power series $(\ref{taylor}$),
which is subject to the inequality ($\ref{raggio}$), where $Q$ is
a function of the electric field $\Ef$.
\section{Conclusions}
In this paper we have proved the existence of high-energy tails
for the EED of the BTE (in the analytic band approximations, and
scattering with acoustic and optical phonon and impurities), with
a moment inequality technique, adapted from the granular flows
\cite{bobylev}.
We have also proved that the 'stretched exponential'
(\ref{f_distr}), used in an heuristic way to model the EED tail,
has a theoretical background.
Our theoretical results are confirmed by MC simulations for bulk
silicon. In particular, in the parabolic band approximation, the
tail is a global maxwellian whose temperature coincides with the
lattice temperature for moderate electric fields whereas, for
higher values, this temperature increases.
In the quasi parabolic case the tail is a 'stretched exponential'
(\ref{f_distr}).
%
%
In the future we shall try  to extent the same technique to other
scattering mechanisms, such as electron-electron.
Another line of research will be to use other analytic techniques
in order to obtain exact information about the order $s$ of the
tail.
%
%
\section*{Acknowledgments}                                                                                               %
%
This work has been supported by MIUR PRIN 2004 "Modelli per il
trasporto di cariche nei semiconduttori: aspetti analitici e
computazionali", "Progetti di ricerca di Ateneo" Universit\`{a}
degli Studi  di Catania, EU Marie Curie RTN project COMSON grant
n. MRNT-CT-2005-019417 .
%
%
%
\newpage
\clearpage
%
%
\begin{figure}
\centering
\includegraphics[width=15cm]{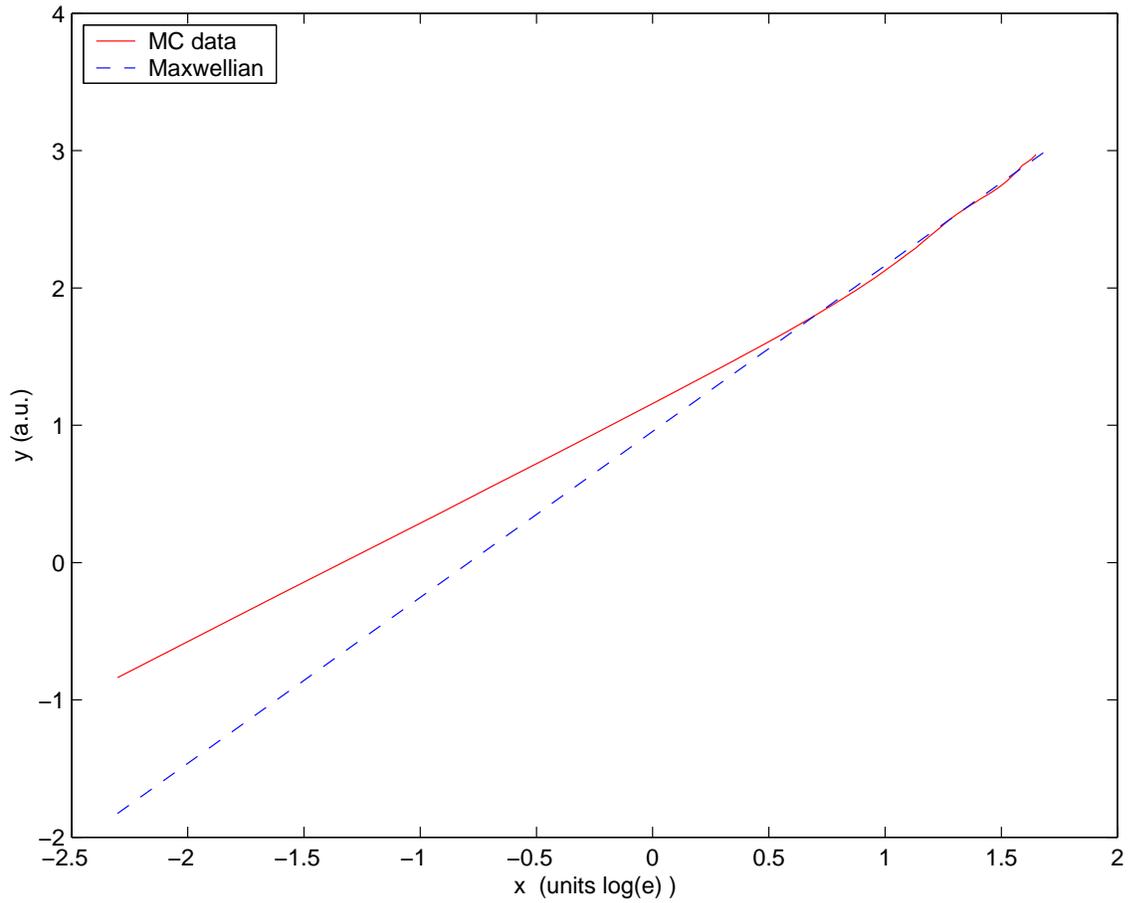}
\caption{Logarithmic plot of the EED obtained with MC simulation
(solid curve) and with eq.(\ref{fit}) (solid dashed straight line)
with $s$=1 and $T_M$ = 4907 K. The electric field is 80 KV/cm  and
the parabolic band approximation is used. \label{fig1}}
\end{figure}
\clearpage
%
%
\begin{figure}
\centering
\includegraphics[width=15cm]{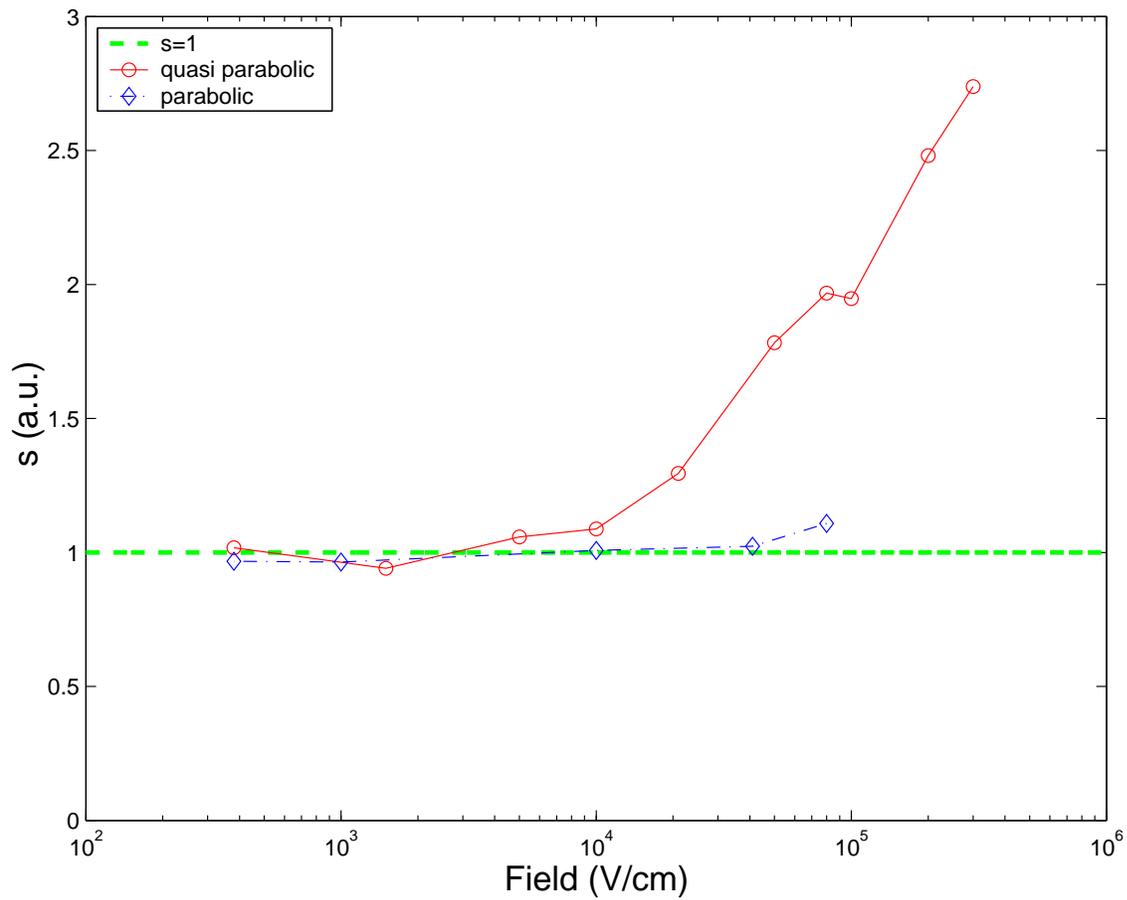}
\caption{The parameter $s$ as function of the electric field, for
parabolic and quasi parabolic band approximations. The dashed
straight line (s=1) indicates a global maxwellian tail.
\label{fig2}}
\end{figure}
\clearpage
\begin{figure}
\centering
\includegraphics[width=15cm]{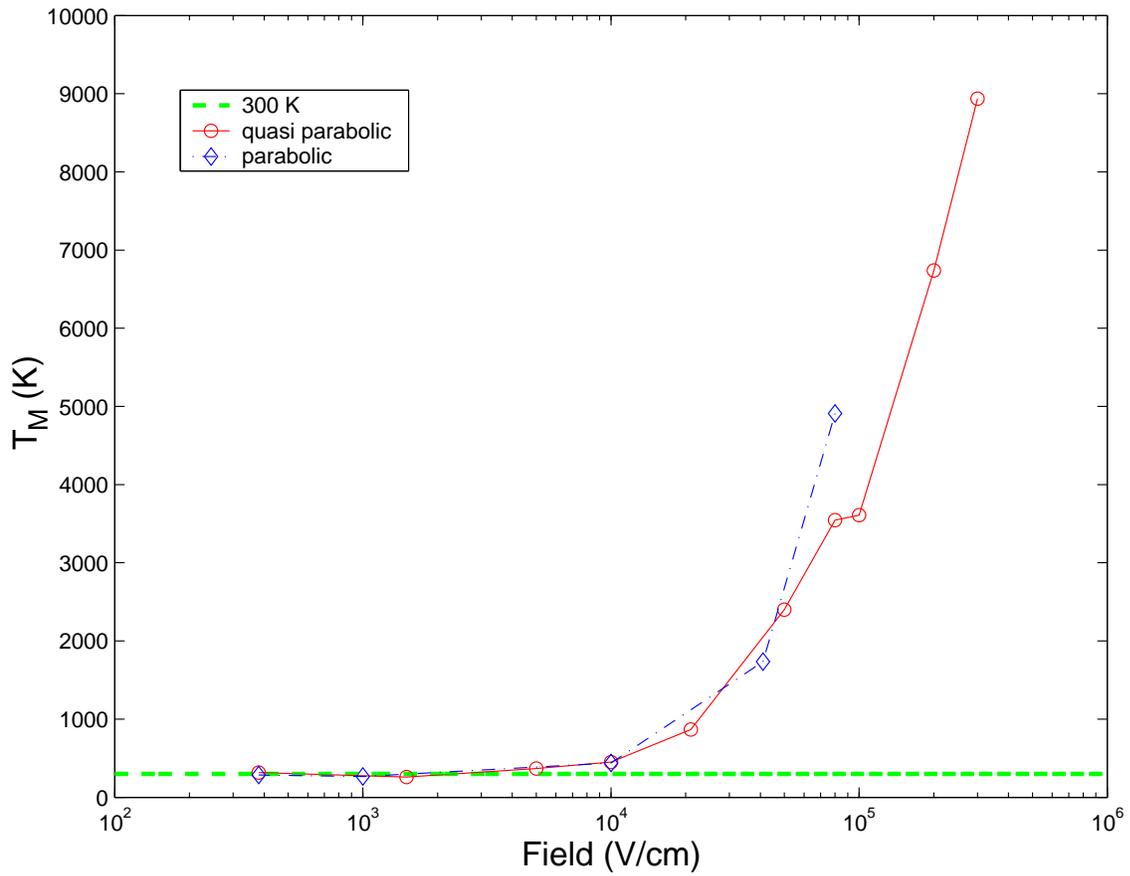}
\caption{The parameter $T_M$ as function of the electric field,
for parabolic and quasi parabolic band approximations. The dashed
straight line indicates the lattice temperature 300 $\mbox{}^0$ K
. \label{fig3}}
\end{figure}
\clearpage
\newpage
%

%
\newpage
%
%
%
%
%
\begin{center}
{\bf Table I. Silicon parameters} \\ [2mm]
\begin{tabular}{|c|c|c|}
\hline
 $ m_e $       & electron rest mass & 9.1095$10^{-28}$ g \\ \hline
 $m^\star$   & effective mass & 0.32 $m_e$ \\ \hline
 $T_L$   & lattice temperature & 300 ${}^{0}K$ \\ \hline
 $\rho_0$  & density &  2.33 g/$cm^3$  \\ \hline
 $v_s$   & longitudinal sound speed  & 9.18 $10^5$ cm/sec \\ \hline
 $\Xi_d$   & acoustic phonon deformation potential & 9 eV \\ \hline
 $ \hbar\omega$ & optical phonon energy & 63 meV \\ \hline
 $ D_t K$ & optical phonon deformation potential & 11 $10^8$ eV/cm \\ \hline
\end{tabular}       \\ [2cm]
\end{center}
\begin{center}
{\bf Appendix} \\ [2mm]
\end{center}
\begin{eqnarray*}
R = \frac{2 \chi}{a\Gamma \beta} \frac{\nq}{\nq +1} \quad , T =
\frac{2^{\frac52}|\Ef|}{\pi(\nq+1)Ba\Gamma \beta} \quad ,
S_2=\frac{2^2}{a \Gamma \beta}\frac{\nq}{\nq + 1}(3\alpha + 8
\alpha^2 \omega)
\end{eqnarray*}
\begin{eqnarray*}
S_1=\frac{2^3}{a \Gamma \beta}\frac{\nq}{\nq + 1}(10\alpha^2
\omega + 9\alpha \omega +1) \quad ,
S_0=\frac{2^5}{a \Gamma \beta}\frac{\nq}{\nq + 1}\omega(2\alpha^2
\omega^2 + 3\alpha \omega +1) \\[1cm]
\end{eqnarray*}
\begin{eqnarray*}
V_3=\frac{2^3}{a \Gamma \beta}\frac{\nq}{\nq + 1}\alpha^2 \quad ,
V_2=\frac{2^3}{a \Gamma \beta}\frac{\nq}{\nq + 1}(8\alpha^2 \omega
+ 3\alpha) \quad ,
\end{eqnarray*}
\begin{eqnarray*}
V_1=\frac{2^3}{a \Gamma \beta}\frac{\nq}{\nq + 1}(10\alpha^2
\omega^2 + 9\alpha \omega +1) \quad ,
V_0=\frac{2^3}{a \Gamma \beta}\frac{\nq}{\nq + 1}(2\alpha \omega^2
+ 3\alpha \omega +1)
\end{eqnarray*}\\[1cm]
where $\chi = 2\alpha^2 + M\Gamma \beta a$ , and $M \in ]0 ,168]$.
\end{document}